# Modelling Charge Coupled Device Readout: Simulation Overview and Early Results

P. Bristow & A. Alexov

*Abstract*

*The Calibration enhancement effort for the Space Telescope Imaging Spectrograph (STIS) aims to improve data calibration via the application of physical modelling techniques.*

*We describe here a model of the Charge Transfer process during read-out of modern Charge Coupled Devices, and its application to data from STIS. The model draws upon previous investigations of this process and, in particular, the trapping and emission model developed by Robert Philbrick of Ball Aerospace.*

*Early comparison to calibration data is encouraging. Essentially, a physical description of the STIS CCD combined with the physics of known defects in the silicon lattice expected to arise in a hostile radiation environment, is enough to yield results which approximately match real data. Uncertainties remain, however, in the details of the model and the physical description of STIS.*

# 1. Introduction

The STIS CCD is known to have a steadily declining Charge Transfer Efficiency (CTE, also hereinafter Charge Transfer Inefficiency, CTI=1-CTE), this is normal for a CCD in a hostile radiation environment. Consequently, charge is lost or deferred as it is read out to the on-chip amplifier from the pixel where it collected during the exposure.

The CCDs in HST instruments present a rather unusual problem. They have a relatively long active lifetime in a hostile environment. The cost of replacing them is prohibitive, yet their CTE is degraded by the continual bombardment of the CCD by energetic particles. As a consequence, HST acquires data from its CCDs which, although of high quality in most respects, suffers relatively large CTI effects. On a ground based astronomical instrument, the CCDs do not degrade in this way. Whilst other applications such as X-ray crystallography, medical imaging and energetic particle detection do sometimes subject detectors to harmful radiation, the detector is easily replaced before radiation induced effects become so large that they need to be corrected. All space based CCD detectors are therefore vulnerable to this degradation in their performance; however few space missions have enjoyed the longevity of HST and have continued to send data after the CTE has degraded significantly.

The effects of poor CTE are easily identified and *qualitatively* understood via consideration of charge trapping and deferral. The principle signature of poor CTE is a position dependent loss of charge (and therefore counts in the readout image) in point like sources. The further the source is from the readout register (see figure 1), the more charge it will lose. We would naturally expect this; a charge packet further from the readout register needs to be transferred more times than one close to the read out register. Further, point like sources, in the presence of significant CTI, will also acquire a tail trailing in the direction in which the CCD rows are read to the register. Once again this effect is stronger the further the source is from the register. This is also easily understood if we consider that the fraction of charge from the original packet which was not transferred effectively may join charge in the packets read out afterwards. Analogous effects are seen, at a much lower level, in the direction of the register read out and can be understood in the same way considering the transfer of charge packets in the register.

Subtler diagnostics of CTI include brighter sources losing relatively more charge and higher background levels reducing the CTI effects across the array. In order to explain these effects we appeal to a slightly more detailed model of the process. The poor CTE is thought to be caused by a population of defects in the silicon lattice which trap electrons (hence we refer to the defects as "traps" hereafter) from the charge packet. Larger charge packets constitute a larger electron cloud which is more likely to interact with the traps. Moreover, if the number of traps is finite, then a high background level means that the traps will easily be saturated. The increase of CTI with time on orbit may also be understood if we postulate that the traps are caused by high energy particles striking the detector.

Clearly then, the level of CTI degradation seen in CCD data is not only a function of the CCD array itself, but also of the illumination pattern in the data in question. Consequently the CTE figure quoted for a device, intended to be the fraction of charge transferred from one pixel to the next in any single transfer, is very much dependent upon the illumination pattern and technique used to derive the quantity. For example, One popular laboratory technique for measuring CTE, known as the Fe55 test, involves using Fe55 radiation to achieve an illumination pattern in which a small fraction of the pixels have a known charge packet whilst the background in the other pixels is low. By comparing the known charge level to those read out from across the chip, it is possible to deduce the charge loss per charge transfer via a simple model. However, if the signal level or background had been different then the CTE value deduced would also have been different. Another way to measure CTE is to read out identical illumination patterns in different directions across the chip. Comparison of signal levels measured for the same source read out in different directions allows computation of a representative CTE value. This technique is suitable for on-orbit calibration of CTE where sparsely populated star fields are usually used. Once again though the value derived in this way is dependent upon the distribution of features in the image, their brightness and that of the background. Measurement of CTE is described further in Appendix B.

The simplest correction applicable to data suffering significant effects of poor CTE is to apply an empirically derived function relating pixel position to lost charge (indeed, this function also needs to relate to epoch and signal level in the pixel and the background). However, in reality charge trapping and deferral lead to more complex effects which can only be well understood by considering the transfer of charge through each pixel and potential trapping site. We discuss here a model of the transfer of every electron through every pixel electrode on the STIS CCD during read out.

The model includes a physical description of the transferral, trapping and emission of charge taken from the CCD literature. This is essentially a set of rules which determine the chances of transfer for each electron. The relevant physical properties of the STIS CCD, such as pixel dimensions, clock read-out timings and operating temperature are also present. In addition the chip is assumed to have a population of charge traps distributed in some way across the array. There are several types of trap, the energy levels and time scales of each are derived from theoretical and laboratory studies whilst their density is related to the time for which the chip has been subjected to radiation damage.

The input to the simulation is a two-dimensional array representing the illumination pattern on the CCD before read out. This may be real STIS data thought to be relatively free of CTI effects or an idealised illumination pattern which may help to understand CTI effects on very specific signal levels and features.

Given this input, the array contents are shifted around in exactly the same way as the contents of the CCD pixels are shifted on the chip during read out. For each and every charge transfer the physical rules are applied. In practice this means that there is a fractional chance of any given electron being transferred, captured, deferred or emitted. At the same time further arrays are used to keep track of the electrons which have been trapped and the status of the various trap types. The simulation and physical models and assumptions are described in some detail in section 2.

At the end of this process, the original illumination pattern has been modified to produce an array of values that would be seen after the read out process, i.e. the simulation has reproduced the CTI effect in the data. There are several ways in which we may assess the accuracy of the simulation. Qualitatively, we may look for well known effects of CTI, by examining the difference between input and output. Quantitatively we may compare to laboratory results for which the charge distribution on a real CCD was known before it was read out, or use on-orbit calibration data which was carefully obtained so that we can infer what the underlying (i.e. unaffected by CTI) illumination pattern was. Early comparisons, of model results to real data, are discussed in section 3.

The model may be fine tuned by tweaking some of the physical parameters; however, we hope to achieve a good fit to the data purely from physical considerations. Ultimately we hope that such a model can be used to correct STIS data. In this case we must take CTI affected data and compute what original illumination pattern would have produced data like this after read out. We discuss this briefly in section 3.

Section 4 discusses our plans for future development . In addition we provide several appendices which summarise other work in this field (upon which this effort draws heavily), describe three popular techniques for measuring CTE and list parameters important to the model.

# 2. The Simulation

## Simulation Overview

The charge distribution after an exposure on the CCD to be simulated is given as an input array. This can be an idealistic or hypothetical image or real data thought to be relatively free from CTE effects (e.g. early STIS data). When using real data it is of course important to remove effects such as bias in order to get back to what the charge distribution on the detector was at the time of read-out. The following discussion relies heavily upon the co-ordinate systems illustrated in figure 1.

We then simply mimic the readout procedure. In terms of array manipulation this means looping over all rows ($j$=1 to $Y$,=1024 for STIS). The contents of the row closest to the register (row 1) are shifted to the register, those of row 2 are moved to row 1, row 3 to row 2 and so on. This effectively means that we are performing a sub loop of $k$=1 to $Y$-$j$ (the –j will become apparent below). Once this has been completed, the register, in reality, is read out, however we are able to treat the register as another 2D array (i.e. we are not limited to just one row of pixels in the register that must itself be read out before the next row can be read in), so we can deal with it separately later. Now $j$ is incremented and the process repeated, except that one row has already made it to the register, so we have $Y$-1 rows left, or more generally $Y$ –$j$. There is, in principle, a further sub loop over each pixel in each row; however, we treat each row as a vector as the pixels are subject to the same operations.

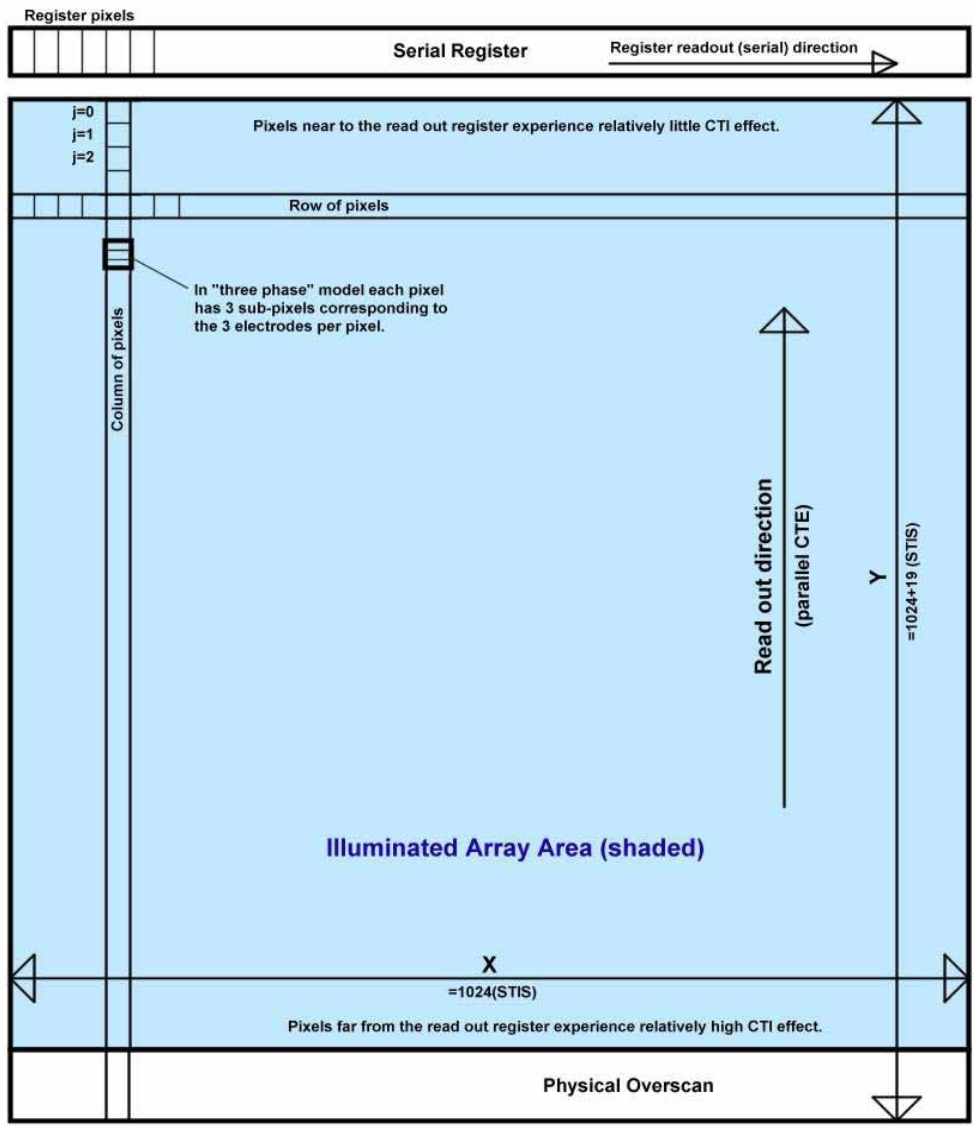

- Figure 1: Schematic of the CCD array.

When each shift is performed the signal held by a pixel may be attenuated by traps capturing electrons or augmented by traps releasing electrons into the pixel. Models for this process are discussed below. Regardless of the process, it is necessary to keep track of the status of the traps as the signal passes through. For this we use further arrays, both for describing trap densities across the chip and for the instantaneous state of these traps.

Once all of the rows have been read into the register "array", the register array is read out row by row as the register on the CCD would have been, except that in reality there is only one row on the real CCD register which has to be read out intermittently with the row shifts from the main array. Clearly there is also only one row of traps to be accounted for in the register, but the status of the traps changes as each row is read out over them, so we cannot treat the columns of the virtual register array as vectors.

The amount of trapping and emission that may occur per shift is limited by the time available. For STIS the serial clocking time is 22μs and the parallel clocking time is 640μs. As the register must be completely clocked out for each parallel shift, the total time between parallel shifts is 640μs+1024x22μs~0.023s. In the more detailed three phase simulation below we consider one phase to include the register readout time and the other two to be only a fraction of the parallel shift time.

At present the output is simply analysed to see how accurately CTE effects measured in real data have been reproduced. However, if this simulation is to be of use in correcting for CTE effects in STIS data then it must be possible to calculate the difference between the STIS raw data and the charge distribution on the chip before it was subject to CTE effects. There are two ways of doing this. Ideally the simulation could be inverted to take the STIS data as input and it would work backwards to calculate the CTE free charge distribution. However, the process is not time-reversible for the Monte Carlo implementation (see below), of course, as a stochastic process it has no unique solution in either direction. Even the non-Monte-Carlo version cannot be easily inverted, the problem of finding the original pixel location of an electron which has been trapped and re-emitted "downstream" is somewhat more difficult than the problem considered here of simulating the trapping and emission process (which is already difficult, and CPU intensive enough!).

The second option would be an iterative or "bootstrap" application of the simulation. The STIS uncorrected data would be the input, but the simulation would be run forwards. The difference between the output and input would then be subtracted from the input and the process repeated. This process would be repeated until the output of the simulation converges to the original STIS uncorrected data. The input should then have converged to the CTE effect free charge distribution. This would only work if the CTE effects were relatively small (which is true).

## Emission/Capture Models

### Ad-hoc Model

To begin with we implemented a model for which the capture probability for each electron in each shift was simply the CTI value. Electrons were re-emitted from traps according to the time constant for the trap types. This model could qualitatively reproduce known phenomena (mitigation of CTE effects by higher background levels, charge trailing bright objects etc.), but was essentially an empirical model in that there were plenty of free parameters which could be adjusted until the results fitted the data.

### Philbrick Model

During pre-launch studies for the Kepler mission at Ball Aerospace, Philbrick developed a readout simulation of this kind with an emission capture model based upon the physics of known bulk state traps (Philbrick 2001). Kepler will use similar CCDs to STIS[1]; therefore Philbrick chose to use STIS as a test case for the model. He found that, using trap densities appropriate to a detector which had been 2.6years in orbit, the results of his simulations showed a global CTE value for STIS which matched the value actually measured after STIS was 2.6 years in orbit.

---
[1] Kepler CCDs will be supplied by both E2V Technologies and STA.

Essentially the model is a simplification of the results of investigations into the behaviour of electron clouds in charge transfer. In describing it here we follow Philbrick (2001) very closely. Fundamental to this model is the idea that as a charge packet passes through it will interact with some fraction of the traps present on the CCD which is dependent upon the size of the charge packet itself. We refer to this fraction as the *exposed traps*, $n_{ex}$, where:

$$n_{ex} = (1 - r_0).n_t.\sqrt{\frac{N_{sig}}{N_{sat}}} + r_0.n_t \qquad (1a)$$

where $n_t$ is the trap density, $r_0$ the fraction of traps exposed when no signal is present, $N_{sig}$ the signal and $N_{sat}$ the saturation signal level. $n_{ex}$ should be evaluated for each trap type modelled. In the presence of a mini-channel (as is the case for STIS), when $N_{sig}$ is less than the min-channel capacity, $N_{sat-mc}$, $n_t$ in equation 1a is replaced with $n_{mc}$, the trap population of the mini-channel, which can be approximated as $n_t L_{mc}/L_{hp}$ (where $L_{mc}$ is the width of the mini-channel and $L_{hp}$ is the horizontal pixel pitch). On the other hand, when $N_{sig} > N_{sat-mc}$:

$$n_{ex} = (1 - r_0).n_{mt}.\sqrt{\frac{N_{sig}}{N_{sat-mc}}} + r_0.n_{mt} + n_{mc} \qquad (1b)$$

where $n_{mt} = n_t - n_{mc}$. Figure 1 in Philbrick (2001) illustrates clearly the dramatic increase in trap exposure that can occur when the signal level increases to just above $N_{sat-mc}$.

The number of empty traps, $n_{empty}$, is simply $n_{ex}$ minus the number of traps known to be filled by the previous charge packet to pass through. The number of traps that become populated, $n_{pop}$, and unpopulated, $n_{unpop}$, is then subject to the following simple rules:

$$n_{pop} = n_{empty}.P_{cn} \; ; \qquad \text{for } N_{sig} \geq n_{empty} \qquad (2a)$$
$$n_{pop} = N_{sig}.P_{cn} \; ; \qquad \text{for } N_{sig} < n_{empty} \qquad (2b)$$

$$n_{unpop} = -n_{empty}.P_{en} \; ; \qquad \text{for storage phase} \qquad (2c)$$
$$n_{unpop} = +n_{filled}.P_{en} \; ; \qquad \text{for barrier phase} \qquad (2d)$$

(Note that $n_{empty}$ can become negative after a large charge packet has passed through a pixel, filling many traps, and is followed by a much smaller charge packet resulting in a much smaller $n_{ex}$.) $P_{cn}$ and $P_{en}$ are the capture and emission probabilities respectively for a dwell period $T_d$:

$$P_{cn} = 1 - e^{-\frac{T_d}{\tau_{cn}}}, \; P_{en} = 1 - e^{-\frac{T_d}{\tau_{en}}} \qquad (3)$$

The trap capture and emission timescales, $\tau_{cn}$ and $\tau_{en}$, reflect the physics of the traps. $\tau_{cn}$ can be calculated from the electron cross section, $\sigma_n$, the thermal velocity, $v_{th}$ (=$[3kT/m^*_{e\text{-}}]$ where $m^*_{e\text{-}}$ is the electron mass) and the signal density (electrons/cm$^3$), $n_s$, via:

$$\tau_{cn} = \frac{1}{\sigma_n v_{th} n_s} \qquad (4)$$

As it turns out, the capture time scales for the parameter ranges that we are interested in are always much smaller than the dwell period, therefore $P_{cn}$ is effectively always unity and capture is assured whenever traps are exposed. $\tau_{en}$ can be calculated from $\sigma_n$, $v_{th}$, the trap energy level $E_T$, the operating temperature, $T$, and the effective density of states in the conduction band, $N_C$ via:

$$\tau_{en} = \frac{\exp\left(\frac{E_T}{kT}\right)}{\sigma_n v_{th} N_C} \qquad (5)$$

(see e.g. Hardy et al 1996, Janesick 2001). Values of $\tau_{en}$ are of the order of the dwell period for some trap types. Note the dependence upon temperature is implicit in the evaluation of both $\tau_{cn}$ and $\tau_{en}$.

In this model three trap types are considered, they are summarized in the table:

| Name | Description | Energy (MeV) | Capture Timescale, $\tau_{cn}$ (μs) at -83°C | Emission Timescale, $\tau_{en}$ (μs) at -83°C |
|---|---|---|---|---|
| P-V (Si-E) | Phosphorus Vacancy Complex | 0.44 | 0.65 | 1.3E+06 |
| O-V (Si-A) | Oxygen Vacancy complex | 0.168 | 0.91 | 0.065 |
| V-V | Divacancy | 0.3 | 0.91 | 80 |

## *One Phase Implementation*

The above emission/capture model was inserted into our simulation code in an attempt to replicate Philbrick's results. Philbrick's own description of the model differs slightly from the above in that he considers even the shifts between electrodes within each pixel. In our initial implementation we simply considered shifts between pixels. The problem with this implementation is that the meaning of barrier and storage phase is less clear. However, careful consideration of the order in which the above rules are applied leads to a model which is a good approximation of the three phase model (below).

## *Three Phase Implementation*

Later we improved the model to consider the shifts between every electrode. Note that this means all of the densities discussed above must be per electrode and the arrays mapping out the trap distribution and states must be per electrode. STIS has a three-phase architecture. The exact clocking of the electrodes is quite complex, so we tried two models (see illustration below) which should represent the maximum and minimum time spent in barrier and storage phases, therefore bracketing reality. The dependence upon the architecture is only slight as can be seen from the Fe$^{55}$ simulation results given below.

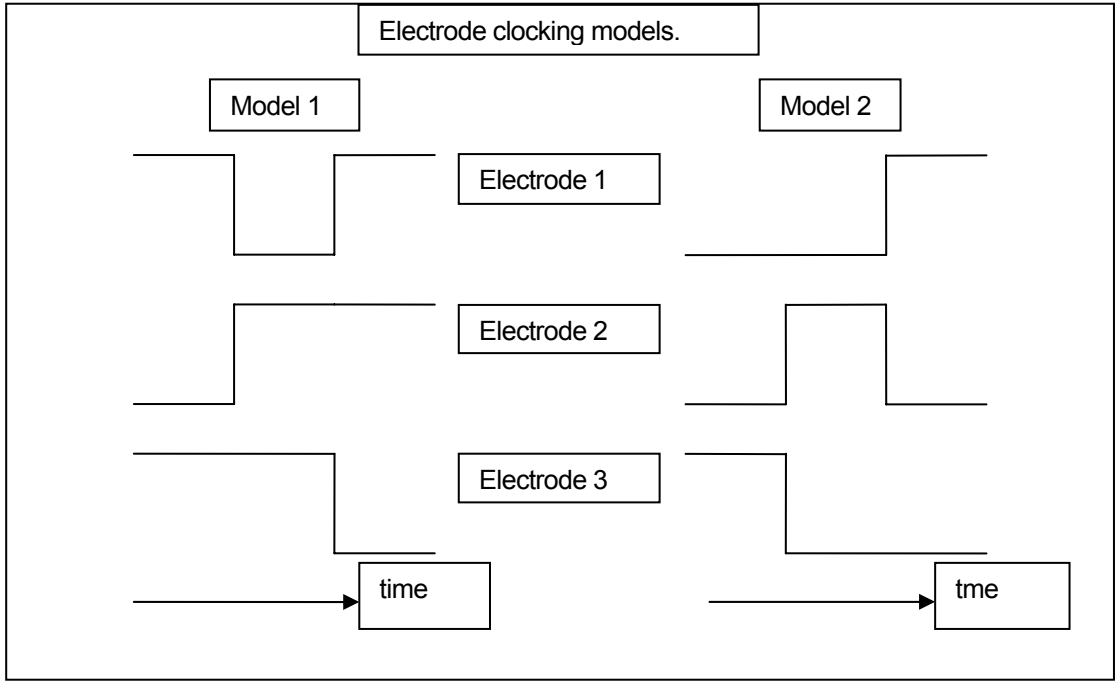

Model 1 was implemented as follows. The array containing the input charge distribution on the CCD was expanded so that every row had two empty rows inserted in between. The traps were also distributed as appropriate across an expanded array. The readout now continued as if there were three times as many rows, however the pixels in the empty rows represented barrier phases, for these pixels only equation 2d was evaluated and the emitted charge was acquired by the nearest "real" row pixel. For pixels in the real rows, equations 2a-c were evaluated. The dwell time was split into one long period (during the readout of the register) and two short periods. In this way every third electrode would have a long period during which a charge packet was present and two short periods without, whilst the two electrodes in between would have only a short period during which a charge packet was present and a short and a long period without.

## *Initial State of Traps*

By default the trap population is assumed to have reached a steady state before readout. This is achieved in the simulation by running the simulation first with all traps empty. Then running the simulation again with the trap population as it was at the end of the first readout and repeating the process a few times until a steady state is reached. This process does not need to be executed for every simulation, it is sufficient to have an appropriate steady state population in a fits file which can be read in at the start of a simulation.

Alternatively the trap population may be set so that all traps are empty or all are full at the beginning of the read-out process. In practise this only affects results significantly for very low signal levels.

## *Integer vs Float Representation of Charge and Traps*

One difficulty in implementing these models was the fact that the charge collected on a CCD is quantized and so too is the number of traps. However, trap densities for the known trap types are typically less than one per pixel, even for CCDs which have suffered significant radiation damage. We have tried two approaches, one in which all quantities are represented by floating point variables and the other in which, though the trap distribution takes fractional values, the number of exposed traps visible at any time is an integer value computed probabilistically as is the quantity of charge trapped therein. The former approach should reproduce global CTE measures accurately, but may be suspect for specific pixel values where the signal is very low.

The latter, Monte-Carlo, approach is much closer to reality, by virtue of the probabilistic nature of the processes. However, it is very slow, essentially the chance of capture/emission for every electron must be considered (in reality this is speeded up a little by considering the chance of *n* from *m* at probability *p*, but even this is highly CPU intensive).

Philbrick's implementation used the floating point approximation (Philbrick, private communication).

## Speed

As the eventual goal of this work is to find a way of predicting corrections to individual datasets, the execution time for the simulation must be reasonable. Unfortunately, for the original implementations coded in Python (with numerical libraries in C), the one phase, floating point implementation required around two hours[2] to simulate the readout of an entire STIS frame, whilst the three phase integer version would have required nearly two days! This is clearly unacceptable especially if several iterations are required in order to reach a solution. One might reflect at this point that we are trying to simulate a readout process which in reality takes around 20 seconds!

In testing we work with just a fraction of the full CCD area. For $Fe^{55}$ simulations a section 128 rows x 64 columns suffice (the result can easily be extrapolated to a higher number of rows). For sparse field tests however it is much less clear how one should extrapolate, therefore we use all 1043 (including over-scan) rows.

The execution time for the parallel readout goes as $Y^2 X^{1/2}$, where *Y* is the number of rows and *X* the number of columns (the power ½ on *X* is an approximation as to reduction in execution time obtained by treating the rows as vectors, without vectorization this would be unity). Therefore, unfortunately, reducing the number of columns used in testing does not help a great deal. The execution time for the serial readout goes as $YX^2$, but this part of the simulation can often be ignored altogether as we are primarily interested in the dominant impact of parallel CTI.

Translating the code to C resulted in dramatically improved performance. The three phase version will now simulate a 1024x1043 STIS readout in around 2 hours and a 64x1043 section in less than ten minutes. However, if this were ever to be considered a pipeline calibration task a further improvement in speed would be required. (Philbrick's implementation was also coded in C, (see Philbrick 2001), he also found the run time prohibitive and generally simulated only sections of the CCD array.)

---

[2] Timings are only approximate and depend upon a number of additional factors. They refer to processing with a Sunfire 280R.

## Distribution of Traps across the Array

The bulk state traps responsible for poor CTE are not likely to be distributed uniformly over every pixel of the array. The distribution will however be random and uniform on larger (~20 pixel?) scales as it is caused by random energetic particle impacts (for STIS there is no reason to believe that any part of the detector surface is more or less protected than any other from such impacts).

Ideally then, to correct data at the individual pixel level for CTE effects we need a map of the bulk traps across the array. This is however very difficult to obtain. The best known method for obtaining such a map is "pocket pumping" (see Janesick 2001). This requires commanding of the CCD clocking voltages so as to allow a low light level flat field charge distribution to be clocked backwards and forwards within the array for many iterations, during which charge will collect in trap concentrations, before being readout. Discrete traps will remove charge causing even single-electron traps to be identifiable when the image is finally read out (see Janesick 2001). To map an entire CCD in this way is a laborious process. Such an experiment was never performed for STIS on the ground and unfortunately there is no possibility to use the on-board STIS commanding to conduct such a test (Goudfrooij private communication).

Another possibility would be to re-analyse ground based $Fe^{55}$ tests conducted on the STIS flight CCD. However this is unlikely to reveal the distribution at the desired resolution and, in any case, the traps responsible for the poor CTE of STIS today, were mostly acquired since launch.

A more speculative idea is the link between the hot pixel distribution (excluding those present at launch) on the detector and the trap distribution. Both features were, after all, created by the same phenomena and they both respond in a similar way to annealing. It would not be difficult to scale the two dimensional trap distributions by the hot pixel map, this remains a future possibility.

Before investigating the effect of the trap distribution at the one pixel level however, we may ask whether the level of clustering amongst the traps may affect the global CTE. We have tried this by increasing the number of traps under randomly chosen electrodes by a fixed factor and reducing the number of traps under all other electrodes such that the total number of traps across the array is not altered.

We found that increasing the clustering increased the global CTE. This can be interpreted as the effect of many traps under one electrode being less than the that of the same number of traps spread over many electrodes. However, one should bear in mind that the trap densities measured in the laboratory were most likely derived with the assumption of a uniform distribution. Therefore if a non-uniform distribution were to be assumed in the laboratory analysis then a different trap density level would have been appropriate and the global CTE effects in this test, by definition, the same.

## Mini-channel

The STIS CCD employs a mini- or "notch" channel (see Janesick 2001, *pp 472*) which is incorporated in this model (see equations 1a & 1b and accompanying discussion). The description of the mini-channel requires two parameters. One is the capacity of the mini-channel, $N_{sat\_mc}$, the other is the mini-channel trap population factor which is the ratio of the mini-channel and horizontal pixel pitches. Neither is very well known for STIS, $N_{sat\_mc}$ being particularly uncertain as many SITe CCDs fabricated in the early 1990s did not have the mini-channel properties they were designed to have and it has even been suggested that STIS has effectively no mini-channel (private communication Rob Philbrick [Ball], Randy Kimbel, [GSFC] and Morley Blouke [SITe]).

This is an unfortunate situation as results display a complex dependence upon the capacity of the mini-channel. When a mini-channel is employed the signal sees significantly fewer traps so long as the signal level is less than the capacity of the mini-channel. However, if the signal exceeds the mini-channel capacity then the signal sees suddenly many more traps, so charge loss increases dramatically. It is especially important to understand this phenomenon when considering adding a "fat zero" or "pre-flash" (see e.g. Cawley et al. 2001 where they discuss this strategy for WFPC2) to data in order to reduce CTE effects. Though the extra background will serve to fill traps which would have otherwise trapped real signal, for certain signal levels, the background added may push the signal above the mini-channel threshold so that many more traps become exposed and the CTE performance will deteriorate.

## Input and Output Formats

The initial illumination pattern can be either an idealised computer generated pattern such as the Fe$^{55}$ test simulation described above, or a Flexible Image Transport System (FITS) file containing real data (assumed to be relatively free from readout effects). FITS data are rotated and clipped so as to obtain the desired array section.

Clearly the ultimate output product is the degraded CTE array which is output as a fits image. In addition the original image section used is output as well as an image showing the difference between input and output so that one sees the net charge loss, or in some cases gain for each pixel. The status of the trap array is also output in FITS format at various stages during the readout process as well as at the end. The record of the final trap array can also be used for subsequent simulations (see the section on initial trap conditions).

More detailed analysis is facilitated by the multivariable plots. The contents of the input and output arrays are output as a simple text list formatted for Xgobi. In this way the input or output pixel values or their differences can be plotted against one another or as a function of row or column. Figure 2 shows data from an internal sparse field test simulation, difference between input and output pixel values is plotted against row number. The rows containing the signal are highlighted. Clearly charge is lost on the forward side of the slit image (yellow) and lost on the trailing side (light blue).

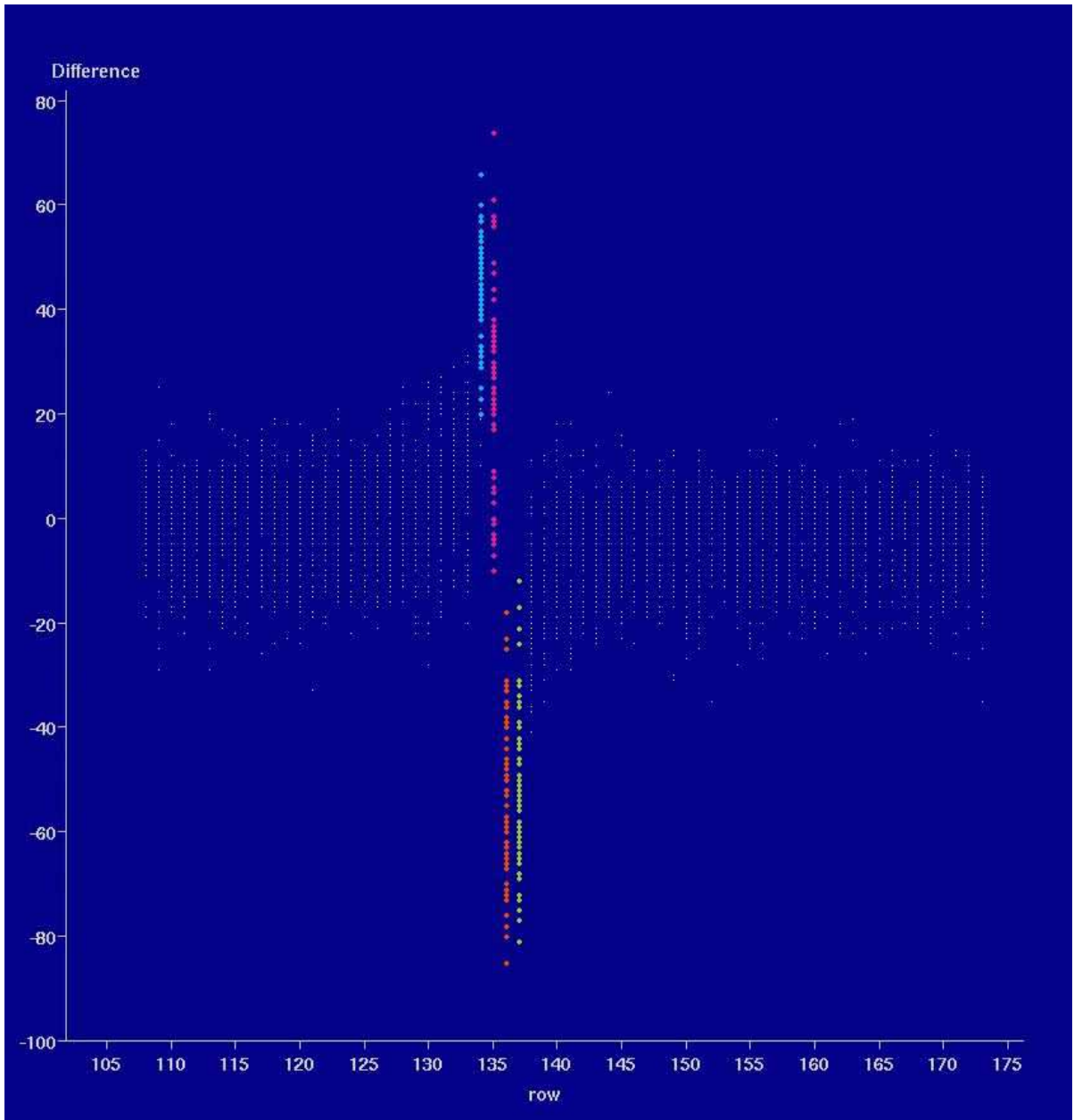

Figure 2: Example xgobi multivariable plot.

An account is also kept of the total number of electrons captured or emitted by the traps under any given electrode. This can be useful for understanding how the trapping actually occurs. See figures 3 and 4 below.

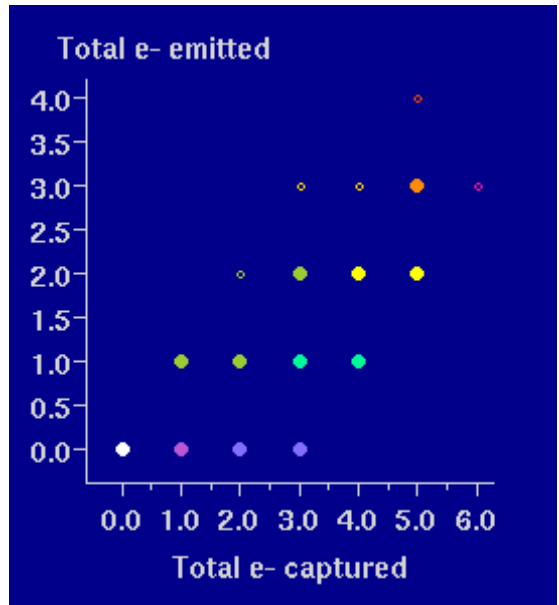
Figure 3: Total e- emitted vs Total e- captured for a simulated Fe55 test. See text for details.

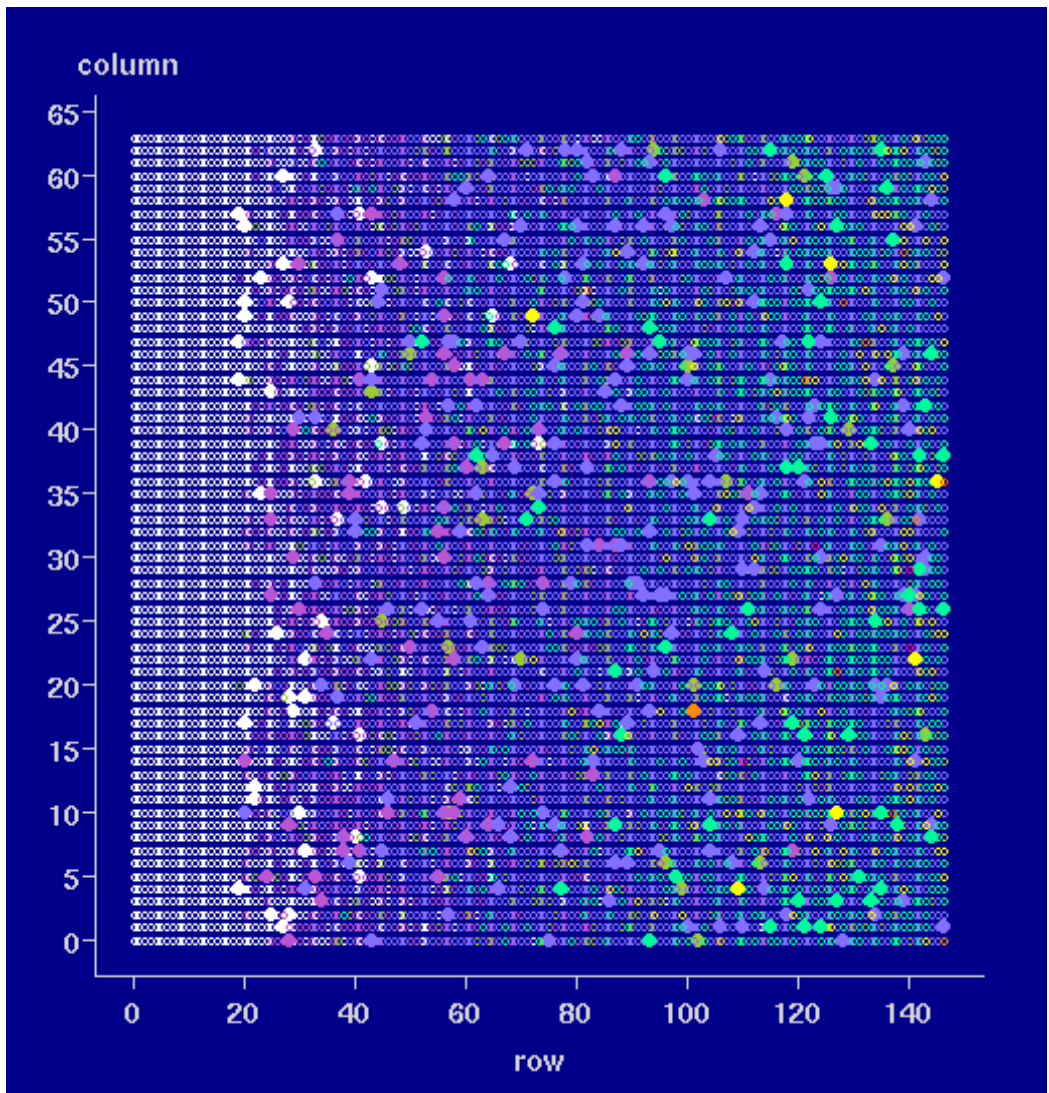

Figure 4: Locations of the traps on the CCD array. Colour coding is as in figure 3. See text for details.

Figures 5 and 6 are two examples of possible input images (5a and 6a) accompanied by images showing the difference between the CTE affected frame that was computed by the simulation and the original input (5b and 6b). The former example is a simulated $Fe^{55}$ test. In the difference image it is clear that the high signal pixels have lost charge (they are dark) which was then re-emitted it the trailing (bright) pixels. The latter is real data. In the difference image we see that the leading side of objects loose charge whilst the trailing side gains.

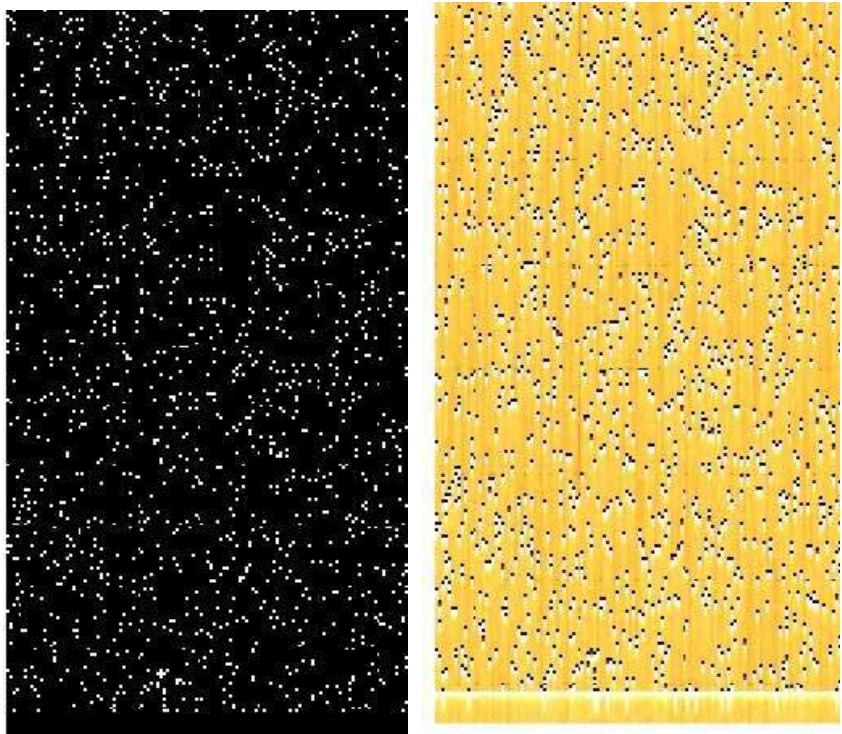

Figure 5: a) Section of Fe55 input data. b) Output difference file.

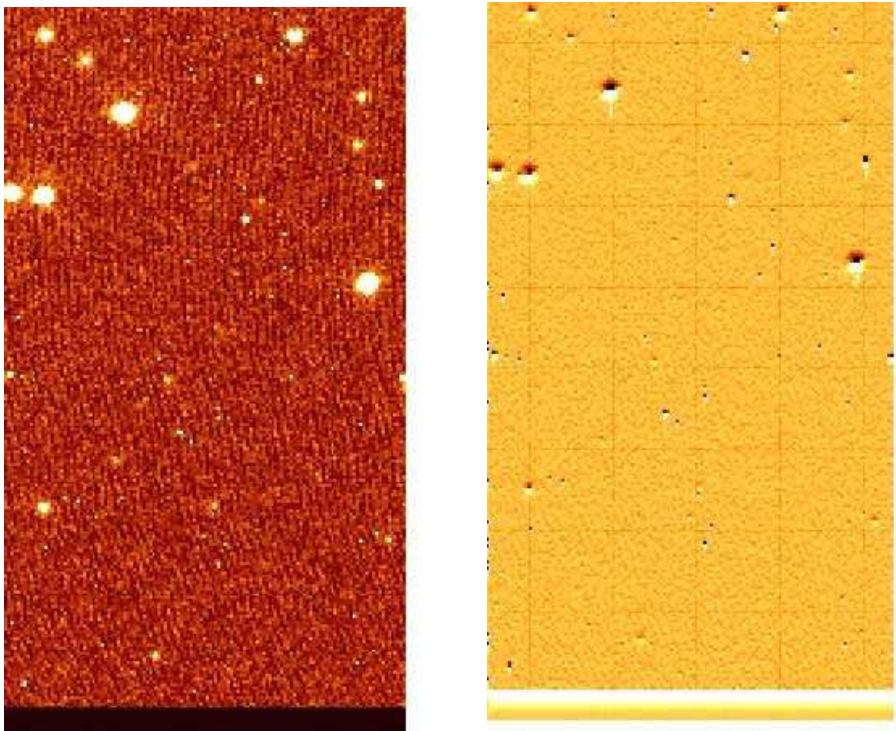

Figure 6: a) Input section of real data. b) Output difference file

# 3. Results

The early testing of the model consisted primarily of simulated $Fe^{55}$ and internal sparse field tests. Qualitative checks were also made on the EPER predicted by the simulation. FITS images of the trap states and CCD array at various stages during output were extensively used during debugging as were multivariable plots of input and output array contents and trap status.

## Sanity Checks

Figure 3 is a plot of the total emitted charge against total captured charge for all pixels on the array. As there were (in this particular simulation) a maximum of three traps under any pixel a fourth electron can only be captured by a pixel whose traps have already emitted at least once, to capture a fifth two must have been released and so on. Note also that in the whole readout (in this case only over 128 rows) there were never more than 6 electrons captured in the traps under any pixel.

Figure 4 shows the positions on the array of the pixels colour coded according to figure 3 (the filled circles are those which received the 1620 electrons from the x-ray source. Clearly the pixels under which more electrons were trapped and released are on the side of the array near to the readout (more charge is read through these pixels).

## $Fe^{55}$ Tests and Trends

An idealised $Fe^{55}$ experiment was simulated to assess how the variation of certain parameters affected the CTE and its dependence upon signal and background. Essentially the input array was simply set up with randomly chosen pixels containing 1620 electrons interspersed with pixels at a given background level. Clearly this is highly idealistic, but it enables a quick estimation of the CTE resulting from a set of input parameters and can be reliably adapted to array sizes smaller than the total (even the Y dimension of figure 1). Moreover, one can easily verify that effects, such as the reduction in CTI for higher background levels, occur as expected.

Figure 7 shows a typical plot from such a simulation which allows the calculation of the apparent CTE. In the data shown, after 128 rows the signal is being readout at an average of 1585 electrons (only the pixels which received 1620 electrons originally are plotted).

$$CTI = 1 - \left(\frac{1585}{1620}\right)^{\frac{1}{128}} \cong 0.000171$$

The table below shows the CTI values derived in this manner from $Fe^{55}$ simulation output for both architecture settings, with and without clustering and with two levels of background. The effects of varying some of the other fundamental parameters are summarised in the table in Appendix A.

| Architecture | Clustering | Background (e-/pixel) | CTI |
|---|---|---|---|
| 1 | Off | 50 | 0.000165 |
| 1 | Off | 10 | 0.000200 |
| 1 | On | 50 | 0.000136 |
| 1 | On | 10 | 0.000141 |
| 2 | Off | 50 | 0.000165 |
| 2 | Off | 10 | 0.000210 |
| 2 | On | 50 | 0.000097 |
| 2 | On | 10 | 0.000107 |

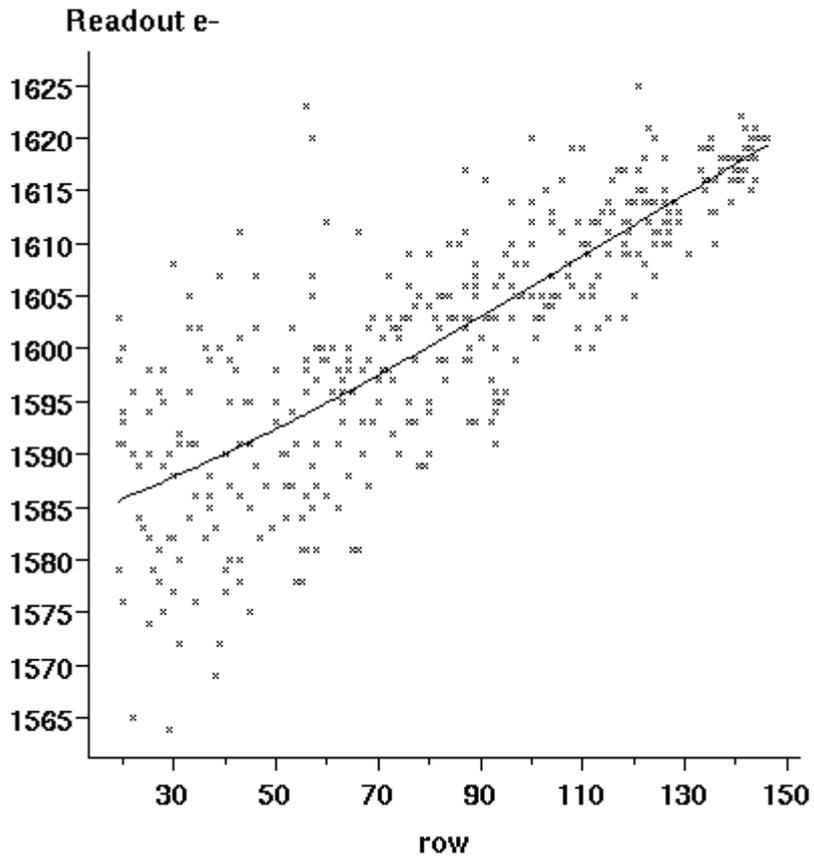

Figure 7: Simulated $Fe^{55}$ test showing readout values of those pixels which originally received 1620e- against row number.

## Comparison to Real data: Internal Sparse Field

We take advantage of the plentiful internal sparse field data available from the STIS archive. These datasets contain slit images of varying intensity and varying distances from the amp used for readout. We choose a dataset in which the slit image is very close to the readout register so that CTE effects are likely to be small (note, this assumption is not critical, all that matters is that we have a slit image to use as our reference, if it has some small distortion to its shape due to CTE effects this will be effectively divided out in the analysis below). We then use this as our input slit image and simulate the read out as if this line was placed at varying distances from the register. The background is also that from the real data.

In this way we are able to produce plots equivalent to those of Kimble et al. (2000). To calculate Kimble et al.'s B amp signal/D amp signal we divide the signal from a read out over $y$ rows by another from a read out over 1024-$y$ rows. Of course our results will necessarily have the property (B signal/D signal)($y$)=1/(B signal/D signal)(1024-$y$) and in particular (B signal/D signal)($y$=512)=1.0. Indeed deviations from this in the real results indicate either a non-uniformity in the trap distribution or a difference in behaviours of the readout amps.

Figures 8 and 9 show our results plotted along with Kimble et al.'s figures 5 a and c respectively. For the lower signal level the agreement is not bad for higher values of y (or "Rows from amp B"), especially given that the loss over 512 rows is almost identical. As discussed above, the data points for $y$=512 and below from Kimble et al. are anomalous. Also in good agreement is the centroid shift for this data, measured at 0.38 pixels in the simulation results compared to ~0.35 (read from the figure 6b of Kimble et al.) for the real data.

For the higher signal level the simulation results appear to underestimate the effect of CTE. The centroid shift in the simulation output is also about 40% lower than in the real data.

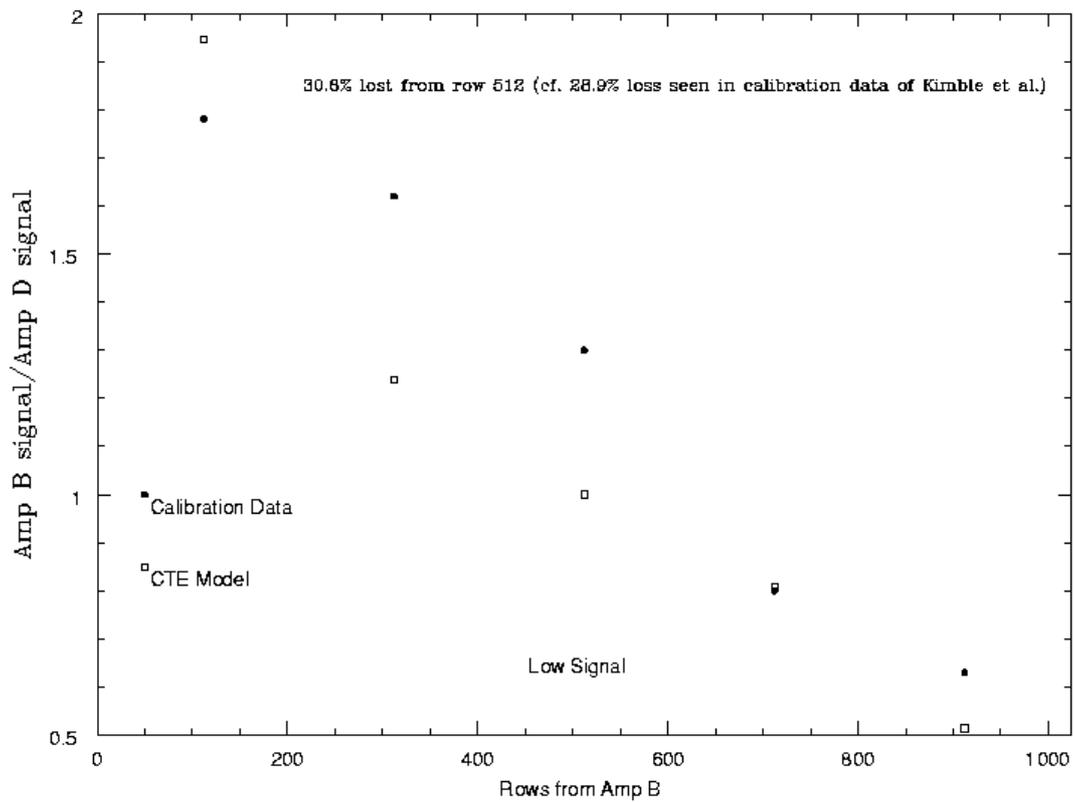

Figure 8: Signal per column 60e- (cf. Kimble et al. 2000, figure 5a)

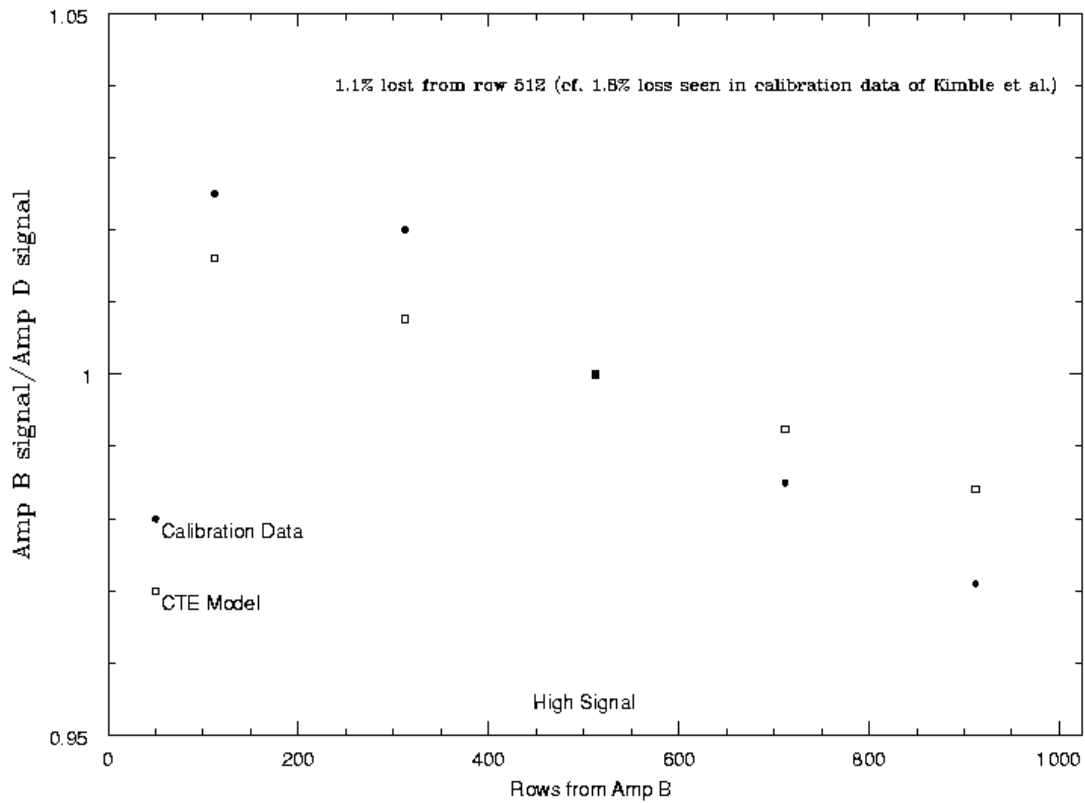

Figure 9: Signal per column 3400e- (cf. Kimble et al. 2000, figure 5c)

A more accurate fit is always possible by arbitrarily fixing trap densities and trapping and emission constants.

A slightly better test would be to compare the simulation output for various values of $y$ to real data for equivalent read-out distances. This turns out to be more difficult as the actual signal received by each exposure varied depending upon the temperature of the calibration lamp. This change in signal received must be factored in before the loss due to CTE can be calculated. Kimble et al. circumvented this problem by carefully analysing the sequence of exposures and interpolating the effective signal. We have not reproduced this analysis yet.

Perhaps a more promising possibility is to look instead at external sparse field data. The many objects in the field of view allow better relative calibration of the two real datasets. An automated object detection algorithm such as SExtractor may be used to find systematic differences in centroid position and intensity independent of readout effects.

Moreover, the physical parameters used are by no means definitive. Uncertainty in the details of the STIS mini-channel is particularly concerning as discussed above.

# 4. The Next Steps

Refinement of the list of physical parameters in the model is a priority. We have a number of leads to follow on this and there is plenty more information available in the literature.

The structure of the model itself has evolved somewhat but is beginning to stabilise. One significant refinement that could be made, and could impact the results, is the following: Each time the trapping and capture is performed in the storage phase, both processes are computed without the knowledge of the outcome of the other. It can occur in the model in its current state that an electron is emitted and trapped in the same cycle, however, what is not taken into account is how the trapping of an electron half way through a phase may affect the probability of the emission of an electron in the remainder of the phase and vice versa. It is hard to predict the impact of this except to say that it would only be relevant for *very* low signal levels, but this could be investigated.

Another feature that could be added to the model is the possibility to replace the uniform trap distribution with a map derived from the STIS hot pixel map (remembering that the hot pixel map is constantly changing, and the appropriate map for a given dataset should be chosen) but scaled so as to have the correct average density.

Regarding testing, the external sparse field test discussed above would seem the most likely investigation to clarify and hopefully resolve the issues raised by the internal sparse field test. However, this would be a time consuming investigation. Another possibility would be to use real $Fe^{55}$ data. If a section of the pre-launch $Fe^{55}$ data, perhaps 50 rows adjacent to the readout register, was used as input for 50 columns in the simulation, then the CTE could be estimated with all of the noise and uncertainty present as in real data. However, there is, of course, no radiation damaged $Fe^{55}$ data from STIS to compare.

Whilst our original intention was to address the more general question of what happens to the illumination pattern from the time when the photon impacts on the detector through to its emergence from the read-out amplifier in ADU form, as is clear from the content of this paper, we have investigated, almost exclusively, the effects of degraded CTE. One concern that we have is that it is not possible to disentangle the effects of charge spreading, CTE and amplifier performance and that a better fit to real data may not be possible without a model that encompasses all of these. Therefore we are also hoping to broaden the scope of the model to encompass other read-out effects.

# 5. Conclusions

We have developed a simulation of the CCD read out process that attempts to reproduce quantitatively the effects of poor CTE in realistic astronomical data. This approach is by no means unique, but we attempt to build upon recent progress in the understanding of the underlying processes. The simulation has been customised to model the behaviour of the STIS CCD, but is in principle portable to other space based instruments. Moreover the model is adaptable to differing operating conditions, illumination patterns and levels of radiation damage.

The simulation gives an insight into the processes involved and serves as a useful tool for better understanding CTE. It reproduces qualitatively, known CTI phenomena. Preliminary quantative results are encouraging. Without any empirical tweaking of the models physical parameters we are able to get an approximate match to various values of CTE measured in orbit. These comparisons now need to be extended to a wider range of data, CTE measurement techniques and radiation damage levels. Simultaneously we intend to improve the physical model by reducing the uncertainty in some parameters and better understanding of some processes. It will be interesting to see if this brings the results into closer agreement with the data.

# Acknowledgements

We would like to thank Rob Philbrick for taking the time to answer our many questions in great detail and Paul Goudfrooij, Randy Kimble, Jim Janesick and Morley Blouke for useful comments.

# Appendix A: Previous Studies

## CCD Literature

The topic of readout from scientific CCDs, and in particular the CTE aspect, has been widely discussed throughout the CCD manufacturer and user communities. However, the seminal reference, to which almost all others refer, is Janesick (2001). Janesick's Chapter 5 covers transfer mechanisms, measurement techniques and the nature of traps in some detail.

In general however, there is surprisingly little discussion of techniques for correcting data obtained with a device which has a poor CTE. This is perhaps because modern devices have excellent CTE and most development effort is directed at improving further the performance of new devices as they reach ever greater chip sizes. As discussed above, HST represents a rather unusual case for which the presence of CTE effects in the data has to be accepted and a way found for correcting for these effects.

Our approach of simulating each and every transfer of charge between electrodes is simple in concept and at least the physics needed to construct such a model is well documented. However there are few examples of other attempts to implement such a model. The most promising existing model for simulating the STIS readout process is that of Philbrick (2002) discussed in detail below. Another model which has influenced this work is that of Hardy et al. (1996, see also Hardy 1997), in which they derive the charge trapped and re-emitted as each charge packet passes, as a function of signal density, from consideration of the clocking phases of a three phase CCD. They do not however simulate each individual charge transfer.

Nevertheless they are able to match experimental data for CTE as a function of temperature and radiation dosage applied to the CCD. Gallagher et al. (1998) arrive at a similar expression for the CTE as a function of signal density and use this in a Monte-Carlo style simulation of the transfer process. In their model they compare a random number (0-1) to the CTE for each electron in every transfer. In this way they are able to reproduce $Fe^{55}$ data (qualitatively) and centroid shifts.

The difference between the above models (and to some extent Philbrick also) and the current work is that their aim was to understand how global CTE properties related to physical properties of traps and radiation dosages. We, however, are more interested in deriving corrections to actual data (in the most optimistic case at the pixel to pixel level) using the physical properties derived by these authors as a basis for our model.

## *Radiation Damage*

In the late eighties and early nineties there was considerable concern about the potential effects of hostile space environments upon CCD detectors. This prompted a great deal of activity at that time in estimating the degradation in performance that may be expected from a given level of exposure to energetic particles.

There are two important types of radiation damage for CCDs; *ionization* and *bulk damage*. Ionization damage arises when energetic particles pass through the sensor's gate insulator causing electron-hole pairs. Consequences include a shift in the clock operating voltages and higher surface dark current generation. Whilst a large shift in operating potential due to ionization damage can degrade CTE performance, this is unlikely to be the case for STIS.

Bulk damage is more likely to be responsible for the gradual decrease in CTE over time seen in data from buried channel devices such as STIS (Hardy 1997). It occurs when incident radiation displaces silicon atoms in the lattice structure creating *bulk traps*. Janesick et al. (1991) demonstrate that these trapping centres can significantly degrade CTE performance. Bulk damage can also lead to hot pixels and dark spikes when electrons from the silicon valence band find their way to trapping centres.

Dale et al. (1993) use the concept of non-ionizing energy loss (NIEL) to evaluate the displacement damage produced in a CCD. NIEL is the energy a particle imparts to a solid through mechanisms other than ionization. To a first approximation, CTE degradation scales with NIEL (Robbins 2000) so that the density of a given type of trap can be predicted as:

$$n_t = K \times \text{NIEL} \times F$$

where $F$ is the particle fluence (per unit area), and $K$ a scaling constant appropriate to the trap type. As the NIEL can be measured experimentally for incident radiation particle types and energy levels, a value appropriate to the operational environment of a space based CCD can be estimated. In this way we can derive $n_t$ as a function of time on orbit.

Gallagher et al. (1998) describe how the experimentally derived bulk trap densities may be used to predict global CTE values via consideration of the Shockley Read Hall theory and a simple Monte-Carlo charge transfer simulation. They are able to reproduce $Fe^{55}$ data from irradiated CCDs.

Hardy (1997) presents a table summarizing the various trap levels reported by investigators. We are most interested here in the P-V, O-V and V-V complexes which are assumed to be the dominant traps in our simulation. A table presenting the parameters used in this work for these traps is given below.

Radiation damage also increases the dark current generation in CCDs via the creation of mid-band traps which enable the transition of electrons and holes between the conduction and valence bands. The increase in dark current is not smooth over the chip but has a grainy structure on a pixel (probably in fact sub-pixel) scale, resulting from the small flaws caused by particle impacts. Three years after launch the number of pixels on the STIS CCD with dark current generation rate greater than 0.01 counts/s was $\sim 2 \times 10^5$.

Radiation damage is, at least in principle, reversible. Heating a CCD to around 150°C (appropriate for annealing the P-V complex, >330°C would be required for the V-V complex) for several hours causes annealing sufficient to recover approximately 80% of the lost CTE (see Robbins 2000 and references therein). Unfortunately this is not an option for STIS. Annealing of hot pixels can however be partially achieved at lower temperatures. Once a month about 80% of the hot-pixels on the STIS CCD are successfully annealed by warming the detector to 0-5°C for 12-24 hours.

Hardy 1997 finds evidence that the trap primarily responsible for increased dark current in irradiated CCDs may not be the P-V complex thought to dominate CTE degradation.

## STIS (at STScI and GSFC)

Monitoring the effects of degraded CTE upon data from all HST instruments has been

 a high priority at STScI and GSFC. Moreover observational techniques and strategies which minimise the likely impact of poor CTE along with post observation corrections have been developed for some instruments. The most comprehensive summary of this work is presented by Cawley et al (2001). They also quantify the impact upon science data by considering the loss in signal-to-noise, brightening of limiting magnitude and brightening of limiting surface brightness as a function of STIS lifetime for several gratings and various source and background conditions. Typically the losses expected are of the order of a fraction of one per cent (of the signal flux) around 2005 and a whole percent around 2010. In terms of limiting magnitudes this corresponds to ~0.1mag (but is highly dependent upon the details of the scenario considered). They go on to show that this will significantly affect the efficiency with which typical STIS science programs may be executed. We note however that this analysis assumes, that some kind of correction is in place, i.e. the photons are accepted to have been lost, resulting in reduced signal to noise, but the knowledge of which photons were lost from which objects is also needed for the data to be used optimally. Cawley et al. also show that the STIS pre-launch CTE was slightly better than that of WFPCII and STIS seems to be degrading at a slightly slower rate.

Bagget et al. (2000) investigate longer timescale traps by examining residual images in WFPC2 data. They conclude that the trapped charge is correlated with the maximum intensity clocked through the pixel. However the level of this effect does not appear to have changed significantly during WFPCII's operational life. This suggests that the traps responsible for residual images are not a product of the NIEL.

Kimble et al. (2000) present a comprehensive analysis of radiation damage effects on the STIS CCD during its first 2.6 years on orbit. Statistical analysis of a very large number of cosmic ray trails (following Riess 1999) shows an asymmetry closely aligned to the parallel readout direction. The average number of electrons in this "parallel tail" after 1024 transfers appears to be increasing linearly with time to around 30 after 2.6years on orbit, whilst a small component of the asymmetry aligned with the vertical (serial) readout direction contains on average less than 5 electrons and a correlation with time is barely detectable. Extended Pixel Edge Repsponse analysis (see appendix B) shows CTI rising linearly with time to around 2.5E-4 after 2.6 years for a signal level of ~100e$^-$/pixel and 1.2E-5 for a signal level of 13,000e-/pixel. Once again, serial CTE effects are very much smaller. The linear relation between CTI and time is exactly as expected due to NIEL in an unchanging radiation environment (no signature of the solar cycle is seen, though with only 3 epochs during a period in which solar activity has peaked and begun to decline we would probably not expect to detect any dependence even if it were present).

Kimble et al. also show results for internal and external sparse field tests which are reasonably consistent, with CTI ranging from ~1.5E-4 for signals of ~500e$^-$/pixel and 2E-5 for signals of ~10,000e$^-$/pixel (after 2.6years on orbit). The STIS CTE effects were also parameterised adopting the formalism pioneered by Stetson (1998) for WFPCII (see also Whitmore et al. 1999 and Dolphin 2000). This parameterisation takes into account sky background and signal level as well as location on the detector and can be calibrated by observing the non-linear nature of source counts vs. exposure time for an object observed on a detector exhibiting poor CTE. The CTI apparent from this parameterisation was lower than the values given above, suggesting that the sparse field tests are "worst cases" (as would be expected).

More subtle effects seen on the sparse field results of Kimble et al. are the row dependent centroid shift of the slit image used for the internal sparse fields and the dependence of the external sparse field CTI results upon the exposure time (in effect the background level). These effects and the dependence upon signal level apparent in the result quoted above demonstrate that CTE effects can only be realistically predicted if the illumination pattern on the CCD is taken into account. Awareness of this prompted investigation at STScI into '…"forward" models that can take a theoretical scene and apply CTE effects to it'. However the conclusion, at least where STIS was concerned, was that known trap levels did not reproduce the observed levels of CTI so this avenue of investigation is currently not being pursued at STScI.

The considerable CTE monitoring data has however been used to derive comprehensive empirical corrections. These will give adjustments to fluxes as a function of location on the detector, epoch, background level and signal level. Separate corrections have been derived for spectroscopic and imaging data Goudfrooij (2002). CTE monitoring for STIS continues at STScI, in particular the acquisition of sparse field data (to add to that already obtained from proposals 7944, 8414, 8910, 8911) necessary to calibrate the empirical corrections.

Hill (2001) obtains results which are somewhat anomalous in that they don't show a linear increase of CTI with time on orbit. He derives the CTI from dark frames using hot pixels as the source of charge on the chip. The resulting plot of CTI versus time shows features that Hill attributes to events in STIS's operational history such as suspends or mechanism resets.

Riess (2000) searches for CTE affects in extended sources and finds evidence of asymmetry in galaxy images which can be attributed to imperfect CTE during readout. He is able to reproduce the effect with a simple readout model similar to that considered here. A similar effect was noted by Freudling (private communication) when investigating data from the STIS parallel survey.

In pre-launch testing Landsman (1996) and Malumuth (1996) both find evidence from sparse field tests for better CTE at the nominal voltage setting than the nominal –3V setting. Malumuth derives values of 0.999995 (five 9s,5) for the nominal setting and four 9s,5 for nominal -3V.

In addition to the continual monitoring of CTE and publishing of empirical corrections, STScI has offered two further options to users concerned that the poor CTE may impact their data. Firstly there is the option for spectroscopic STIS data that the slit image is placed upon rows close to the readout register. Secondly users of WFPCII may opt to have a "fat zero" (additional flat background) added to their data. This has the penalty of reduced signal-to-noise, but the CTE degradation should be reduced (but see the "mini-channel" discussion in section 2).

# Appendix B: Measures of CTE

*It is not possible to quote a single figure for CTE that covers all operating conditions, image types and CCD structures*. Measured CTE figures are only directly relevant to the device type under test, running under the test operating conditions and with the images used for the test measurement. We give here a brief overview of some of the CTE tests relevant to STIS: consult the references given for a more thorough description.

## $Fe^{55}$

The $Fe^{55}$ soft x-ray technique (pp. *418* Janesick 2001) has become a standard for measuring CTE in the laboratory. $Fe^{55}$ atoms decay into Mn atoms when a K-shell electron is absorbed by the nucleus. An x-ray is generated when an electron drops from one of the outer shells to fill the vacancy left in the K-shell. The physics describing what happens when such an x-ray is absorbed by the CCD is quite involved (see pp.*132* of Janesick 2001 for the details), but the important result is that signals are be produced on the CCD array (usually confined to just one pixel) at five different well defined levels, one of which is Kα ($1620e^-$). Hence the signal in the pixels before readout is known with some certainty. After readout a plot can be made of signal vs. distance from readout serial register (for parallel CTE). Pixels close to the register will have values close to $1620e^-$ (plus background). Those more distant will have had there values attenuated by poor CTE, the gradient of the plot can be used to estimate CTE.

$Fe^{55}$ is a fairly robust and repeatable technique. However space based missions could not carry the necessary equipment to monitor the detector degradation while in orbit via this technique. Another drawback is that it measures the CTE for a rather extreme illumination pattern on the array, that is, relatively low background between signals which are confined to just one pixel.

Nevertheless, the concept provides a convenient test for our simulations. It is quite simple to set up an input array image which consists of randomly chosen pixels containing 1620 counts interspersed with pixels at a given background level. Clearly this is highly idealistic, but it enables a quick estimation of the CTE resulting from a set of input parameters and can be reliably adapted to array sizes smaller than the total (even the Y dimension of figure 1). Moreover, one can easily verify that effects, such as the reduction in CTI for higher background levels, occur as expected.

## EPER

Extended Pixel Edge Response is much more easily monitored during space missions as it simply requires a flat-field exposure. CTE is then inferred from the charge in the physical over-scan region which was not illuminated during the exposure (see figure 1). It is defined as:

$$\text{CTE}_{\text{EPER}} = \frac{S_{\text{EP}}}{S_{\text{LP}} Y}$$

where SEP is the total signal in the register and SLP is the signal in the last pixel of the exposed array. The last pixel is specified because, although a flat-field was used, charge from neutral material outside the array diffuses into the last pixel so that it often contains more charge than others in the column. This, unfortunately, makes this test rather hard to reproduce in the simulation as there is no model to describe this diffusion effect. Moreover, EPER results are also sensitive to the behaviour of the readout amp. The STIS readout amps are known to respond in a non-linear fashion to sudden changes in signal such as occur at the transition between the main array and the over-scan region. Therefore no quantitative comparisons have been made between simulation output and STIS EPER data.

## Sparse Field

This STIS version of this technique was developed by the STIS team (Kimble et al. 2000) and (in the case of parallel CTE) makes use of the ability of the STIS CCD to read out to registers on opposite sides of the array (for serial CTE the same method makes use of amplifiers at opposite ends of the registers). A sequence of nominally identical exposures is taken, alternating the readout between the registers. The observed differences between the results obtained from the two registers can be fit to a simple CTE model.

Calibration data have been obtained at STScI for two tests based upon this idea. The "internal sparse field test" uses slit images from a flat field lamp placed at various positions across the detector and read out through opposite registers. The ratio of the line of the signal readout from the slit image from the two amplifiers allows the calculation of the CTE (Kimble et al 2001). The "external sparse field test" makes use of exposures of the outer regions of globular clusters, once again read out through opposite registers. This has the advantage that, even though there is always some level of uncertainty in HST's pointing, statistical analysis of the many objects in the field of view allows the two datasets to be reliably compared.

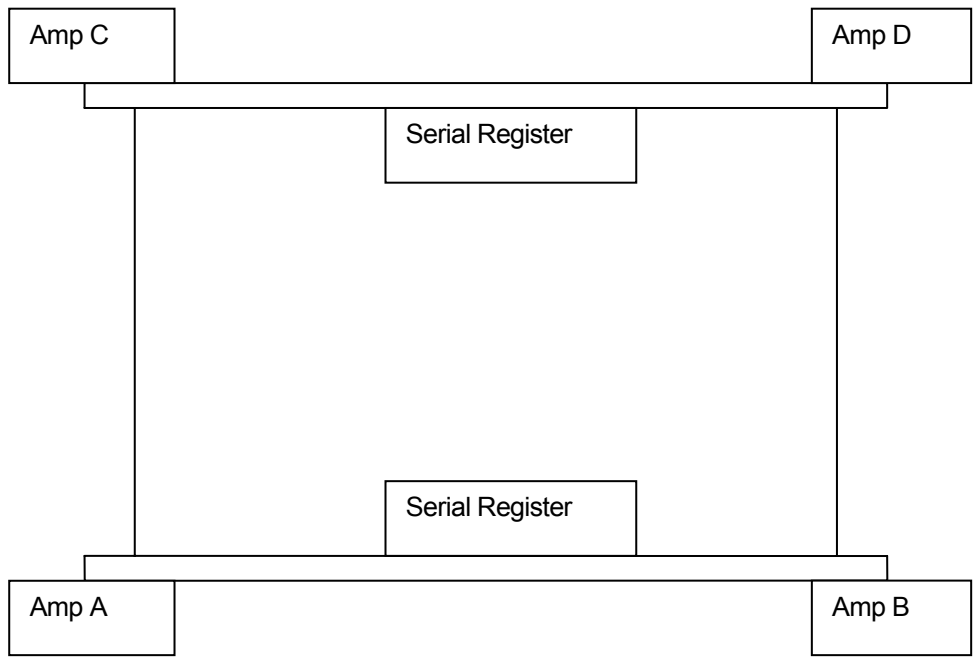

The STIS dual register and readout amplifier layout.

# Appendix C: Summary of Selected Model Settings

| Parameter | Comments | Possible Values | Effect | Sources |
|---|---|---|---|---|
| $r_o$ | Fraction of traps exposed at zero signal (equation 1) | 0.28 | Higher $r_o$ => lower CTE at low signal levels. | Determined empirically, see Philbrick (2002) |
| $n_t$ | Density of bulk state traps, function of trap type and NIEL (equation 1) | In the range $0.1 < n_t < 10$ for all trap types modelled here. | Higher $n_t$ => lower CTE | Determined experimentally, e.g. Robbins (2000) |
| Trap clustering | Currently only a test implementation, appears to be significant | On/off | Trap clustering="on" results in higher CTE for a given $n_t$ | None |
| Initial state of traps | Probably should assume that traps have reached equilibrium before readout begins. | Empty, full, equilibrium | Empty initial state => lower CTE, however the effect is marginal and less for pixels further from the readout Amp. | None |
| Capture time constants, $\tau_{cp}$ | Depends on trap type, however these values are always so very much lower than the dwell period that capture is virtually assured. | $10\text{ns} < t_c < 10\mu\text{s}$ for all trap types | Little effect as realistic values are always so low that $P_c$ is effectively always unity. | Philbrick (2000 + private communication) |
| Emission time constants, $\tau_{em}$ | Depends on trap type, in some cases is comparable to the dwell time. | $10\text{ns} < t_e < 10\text{s}$, with PV traps much longer than the others. | Obviously effects the re-emission of trapped charge (scale length of high signal trails). Also effects when traps become empty and ready to trap again. | Philbrick (2000 + private communication) |
| Mini-channel capacity $n_{\text{sat-mc}}$ | Poorly known, see section 4 | ~2000e-, though there seems to be some uncertainty as to the actual value. | Complex, see discussion | Philbrick, private communication |

# Appendix D: Core Physical Parameters

| Parameter | Comment | Value(s) | Units |
|---|---|---|---|
| $n_t$(O-V) | Density of O-V traps for uniform distribution | **0.341**; 0.365 | Traps per electrode |
| $n_t$(P-V) | Density of P-V traps for uniform distribution | **0.455**; 0.642 | Traps per electrode |
| $n_t$(V-V) | Density of V-V traps for uniform distribution | **0.228**; 3.68 | Traps per electrode |
| $n_{mc}$ factor | Factor to give density of traps in the min-channel | 0.185 | - |
| $N_{sat}$ | Saturation level for pixel | 144,000 | e- |
| $N_{sat-mc}$ | Saturation level for mini-channel | 0; **2,000**; 15,000 | e- |
| $\tau_{cp}$(O-V) | Capture time constant for O-V trap | 0.911; **0.134** | µs |
| $\tau_{cp}$(P-V) | Capture time constant for P-V trap | 0.911; **0.047** | µs |
| $\tau_{cp}$(V-V) | Capture time constant for V-V trap | 0.650; **0.157** | µs |
| $\tau_{em}$(O-V) | Emission time constant for O-V trap | **0.228**; 0.065 | µs |
| $\tau_{em}$(P-V) | Emission time constant for P-V trap | **2.68**; 1.3 | S |
| $\tau_{em}$(V-V) | Emission time constant for V-V trap | **725.**; 80. | µs |

Where more than one possible value are given this reflects more than one possible estimate from the literature. The values in bold type face are the ones used for all results presented here.

The trap density values should correspond to a Total NIEL on the STIS detector 1.29E+07MeV/g, which is the estimate for 2.6years on orbit (Philbrick 2001)

# References

Bagget, S., Birreta, J. & Hsu, J.C., ISR WFPC2 **00-03**, 18[th] September **2000**

Banghart, E. K., et al., IEEE Transactions on Electron. Devices., Vol. **38**, No. **5**, pp. *1162*. **1991**

Cawley, L., Goudfrooij, P., Whitmore, B., ISR WFC3 2001-05, **2001**

Dale et al. IEEE Transactions on Nuclear Science, Vol **40**, No. **6**, pp. *1628*, **1993**

Dolphin, A. E., PASP, **112**, *1397*. **2000**

Gallagher, D et al., SPIE Vol. **3301**, pp. *80.* **1998**

Goudfrooij, P. STIS ISR No. **?**, in preparation, **2002.**

Hardy, T. D., M.Sc. Thesis, Simon Fraser University, **1997**

Hardy, T., D., M.J., Murowinnski, R., "Optical Detectors for Astronomy", Eds Beletic, J. W. & Amico, P., **1996**

Hill, R. S., STIS ISR GSFC **64**, 15 March **2001**

Janesick, J., "Scientific Charge Coupled Devices", SPIE, ISBN 0-8194-3698-4, pp. *387,* **2001**

Janesick et al., pp. *87*, SPIE Vol **1447** Charge-Coupled Devices and Solid State Optical Sensors II. **1991**

Kimble, R. A., Goudfrooij, P. and Gililand, R.L., Proc. SPIE Vol. **4013**, p. *532-544*, UV, *Optical, and IR Space Telescopes and Instruments*, James B. Breckinridge; Peter Jakobsen; Eds. **2000**

Landsman, W. STIS Pre-launch TR (GSFC) No. **52**, 31[st] July **2000**

Malumuth, E, STIS Pre-launch TR (GSFC) No. **55**, 14[th] August **2000**

Philbrick, R. H., "Modelling the Impact of Pre-flushing on CTE in Proton Irradiated CCD-based Detectors", Ball Aerospace & Technologies Corp. **2001**.

Robbins, M. "The radiation Damage Performance of Marconi CCDs", Marconi Applied Technologies, Technical Note: *S&C906/424,* 17 Feb **2000** (available on request)

Riess, A., WFPCII ISR 99-04. **1999**

Riess, A., WFPCII ISR 00-04. **2000**

Stetson, P. B., PASP **110**, *1448*. **1998**

Whitmore, B., Heyer, I. & Casertano, S., PASP **111**, *1159*. **1999**